\documentclass{nature}
\usepackage{eqnarray}
\usepackage[english]{babel} 
\usepackage[]{layout}
\usepackage{amsmath,amsfonts,amssymb}

\usepackage{graphicx}
\usepackage{float}
\restylefloat{table}
\usepackage{bm}
\usepackage{color}
\usepackage{braket}
\usepackage{array,multirow,makecell,multicol}
\usepackage[markup=default]{changes}
\usepackage{multirow}
\makeatletter
\let\saved@includegraphics\includegraphics
\AtBeginDocument{\let\includegraphics\saved@includegraphics}
\renewenvironment*{figure}{\@float{figure}}{\end@float}
\makeatother
\usepackage{diagbox}
\usepackage{tabularray}

\title{Spin-to-orbital angular momentum conversion in non-Hermitian photonic graphene}

\author{Zhaoyang Zhang$^{1\ast\dag}$, Pavel Kokhanchik$^{2\dag}$, Zhenzhi Liu$^1$, Yutong Shen$^1$, Fu Liu$^1$, Maochang Liu$^3$, Yanpeng Zhang$^1$, Min Xiao$^{4}$, Guillaume Malpuech$^{2\ast}$, Dmitry Solnyshkov$^{2,5\ast}$}

\begin{document}
\maketitle
 
\begin{affiliations}
\item Key Laboratory for Physical Electronics and Devices of the Ministry of Education \& Shaanxi Key Lab of Information Photonic Technique, School of Electronic Science and Engineering, Faculty of Electronics and Information, Xi'an Jiaotong University, Xi'an 710049, China
\item Institut Pascal, PHOTON-N2, Universit\'e Clermont Auvergne, CNRS, Clermont INP,  F-63000 Clermont-Ferrand, France
\item International Research Center for Renewable Energy \& State Key Laboratory of Multiphase Flow in Power Engineering, Xi’an Jiaotong University, Xi’an 710049, China
\item National Laboratory of Solid State Microstructures and School of Physics, Nanjing University, Nanjing 210093, China
\item Institut Universitaire de France (IUF), 75231 Paris, France
\item[] $\dag$ These authors have contributed equally to the work.
\end{affiliations}

\begin{abstract}
Optical beams with orbital angular momentum (OAM) have numerous potential applications, but the means used for their generation often lack crucial on-demand control. In this work, we present a mechanism of converting spin angular momentum (SAM) to OAM in a non-structured beam. The conversion occurs through spin-orbit coupling in a reconfigurable photonic honeycomb lattice with  staggering implemented by electromagnetically-induced transparency in an atomic vapor cell.
The spin-orbit coupling allows to outcouple the OAM signal from a particular band in a given valley determined by the chirality of  light or the lattice staggering, providing a non-zero Berry curvature for generating OAM.
The dependence of the output OAM on the chirality of the input beam  is the first control knob. The staggering works as a second control knob, flipping the sign of OAM for the fixed chirality. The demonstrated conversion between SAM and OAM is important for optical communications. Our results can be extended to other implementations of paraxial photonic graphene.
\end{abstract}

\maketitle

The quest for efficient information transfer makes scientists look for new degrees of freedom. In photonics, the orbital angular momentum (OAM) of light beams~\cite{yao2011orbital,bozinovic2013terabit,willner2015optical,miao2016orbital} and individual or entangled photons~\cite{vallone2014free} is now seen as a very promising candidate~\cite{erhard2018twisted} for data encoding thanks to its unbounded nature~\cite{sit2017high}. Entanglement has recently been demonstrated for the OAM degree of freedom~\cite{huang2025integrated}.
Non-zero OAM beams are also intensively applied to create tractor beams~\cite{gao2017optical} and holographic optical tweezers~\cite{grier2003revolution,padgett2011tweezers}. 

Various techniques are currently used for OAM generation and control in coherent light beams \cite{ni2021multidimensional}, with the shaping of paraxial beams by spatial light modulator being the most widespread and well-established technique~\cite{wang2012terabit,bozinovic2013terabit}. 
Berry curvature can be exploited for the  generation of OAM from topological interfaces in quantum Hall regime~\cite{bahari2021photonic} because of its inherent relation to angular momentum~\cite{Chang2008}. Other topological structures are also used~\cite{hwang2024vortex,hu2024topological}.
Conversion of the photonic spin angular momentum (SAM), also known as chirality or circular polarization degree, to OAM \cite{brasselet2011electrically} relies on complex architectures of metasurfaces~\cite{piccardo2022vortex}, q-, J-, and other waveplates~\cite{marrucci2006optical,naidoo2016controlled,Rego2019}, 
or spin-orbit coupling (SOC) in sophisticated semiconductor micro-structures~\cite{carlon2019optically}. Berry curvature associated with SOC is also employed to realize SAM to OAM conversion \cite{Biener2002,marrucci2006optical,karimi2014generating}. Though the methods are numerous, the on-demand control of the conversion is rarely achieved (see \cite{mancini2024multiplication} for an example).

All mentioned systems with SAM to OAM conversion have SOC as a common component, but not every type of SOC is able to produce this conversion. 
In general, photonic SOC can have various forms producing fascinating effects, such as the spin-Hall effect of light~\cite{Onoda2004} or optical spin-Hall effect~\cite{Kavokin2005}, but with an important common feature: it is always related to the transverse nature of the light waves~\cite{Bliokh2008}. Any spatial inhomogeneity allowing to define the transverse-electric (TE) and transverse-magnetic (TM) directions, necessarily gives rise to a TE-TM SOC. So, every photonic lattice naturally possesses it. 

Photonic graphene based on electromagnetically-induced transparency (EIT) has recently emerged as a versatile reconfigurable platform with suitable SOC allowing OAM generation~\cite{Zhang2019,Zhang2020,Zhang2022}. Other types of photonic graphene are also actively studied~\cite{Rechtsman2013b,polini2013artificial}. The valleys of the photonic graphene are characterized by Berry curvature~\cite{Ozawa2019}. The generation of an OAM quantum, that is, an optical singularity called a quantum vortex, requires exciting a given valley (Dirac point) of the Brillouin zone of photonic graphene, but with a structured beam covering only A or B sites of the unit cell~\cite{song2015unveiling,Zhang2019}. Changing the sign of the generated OAM requires either switching the excited valley from $K$ to $K'$ (changing the incident direction of the probe beam), or switching the type of the excited sites (A to B) by rotating the lattice, which in both cases is a macroscopic modification. An additional tunability can be provided by staggering, namely, the energy difference (gap) between the A and B sites of the unit cell. Staggered photonic graphene allows to generate OAM either by exciting a given valley with a spatially homogeneous probe, which requires a fine tuning of energy to the upper or lower band in the valley, or by switching the valley, which again are macroscopic modifications of the excitation. And finally, all described regimes show OAM generation, but not SAM to OAM conversion.

In this work, we demonstrate experimentally SAM to OAM conversion of a non-structured probe beam in a staggered photonic graphene and its control by the  staggering without modifying the excitation scheme. We derive theoretically the SOC term in non-Hermitian staggered photonic graphene and explain how it provides the polarization conversion of the OAM signal which accumulates thanks to Berry curvature. This mechanism of SAM to OAM conversion can be realized in other paraxial implementations of photonic graphene \cite{Peleg2007,Rechtsman2013b,polini2013artificial,Plotnik2014}
and can open large possibilities for applications in OAM generation for optical manipulation \cite{gao2017optical,grier2003revolution,Arikawa2020} and microscopy \cite{willig2006sted,Lavery2013}, and in classical and quantum communications \cite{naidoo2016controlled,fickler2012quantum}, including integrated photonic devices~\cite{Chen2020}.

\section*{Results}

\begin{figure}[tbp]
\centering
\includegraphics[width=0.9\linewidth]{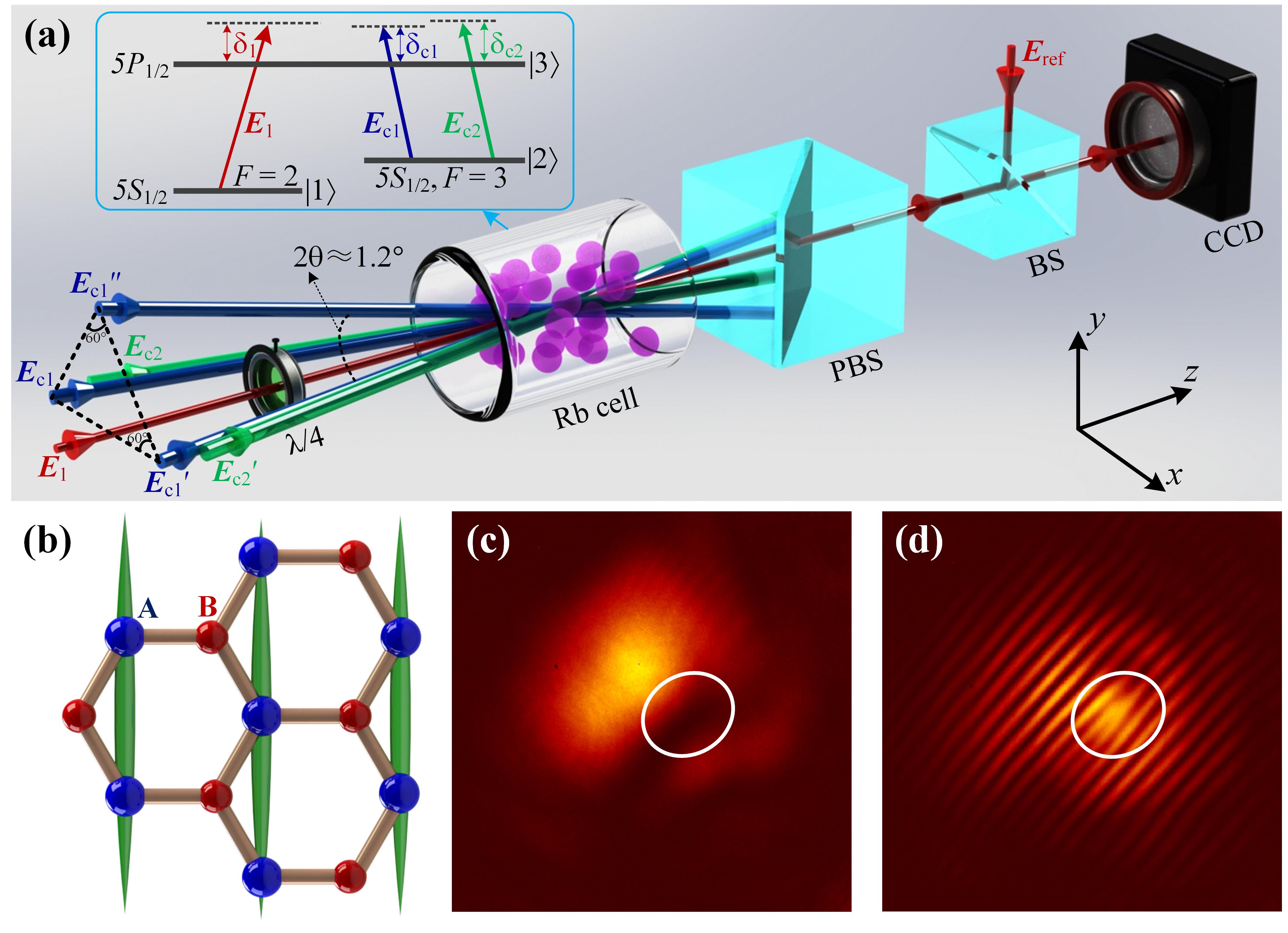}
\caption{\small \textbf{Photonic graphene in a Rb cell: scheme of the experiment and detection of OAM.}
\textbf{(a)}  Experimental setup. Three coupling beams $\textbf{\emph{E}}_{c1},\textbf{\emph{E}}_{c1}',\textbf{\emph{E}}_{c1}''$ (blue arrows) with an angle of $\sim1.2^\circ$ between each two of them form a hexagonal interference pattern; two additional coupling beams $\textbf{\emph{E}}_{c2}, \textbf{\emph{E}}_{c2}'$ (green arrows) create 1D interference fringes. While passing through Rb vapor cell with the EIT configuration, these five coupling beams induce a staggered honeycomb photonic potential for the probe beam $\textbf{\emph{E}}_1$ (red arrow). A polarizing beam splitter (PBS) cube filters out $y$-polarized coupling beams, and only the $x$ component of the probe beam is detected by the CCD camera. Inset: Atomic energy-level structure; $\delta_1$ ($\delta_{c1}$, $\delta_{c2}$) is the frequency detuning of the field $\bm{E}_1$ ($\bm{E}_{c1}$, $\bm{E}_{c2}$).
\textbf{(b)} The schematic of the formed honeycomb photonic lattice with staggering due to the 1D interference fringes (green lines) covering the A sublattice (blue dots).
\textbf{(c)} Observed output probe beam with a circular-polarized input. \textbf{(d)} The interference between the output probe in (c) and the reference beam $\bm{E}_{ref}$. }
\label{fig1}
\end{figure}

We create a staggered honeycomb photonic potential under the EIT regime in a Rb vapor cell. The potential varies in the $x\text{-}y$ plane while it is translationally invariant along the $z$ axis inside the cell. Figure~\ref{fig1}a shows the experimental setup. The honeycomb photonic lattice is created by the interference of three coupling laser beams (blue arrows in Fig.~\ref{fig1}a). Interferometric fringes of two extra coupling beams (green color in Fig. 1a,b) cover one set of lattice sites to introduce the staggering by modifying the refractive index. We explore the dynamics of the system near the $K$ point by sending a probe beam (red arrow in Fig.~\ref{fig1}a). Since we are interested in studying SAM to OAM conversion, the probe beam is always circularly polarized (right or left) in our experiments. The probe experiences effective susceptibility depending on the parameters of coupling beams (see Eq.~\ref{EIT} in Methods and Section 1 of Supplementary Materials), creating the lattice and staggering (Fig.~\ref{fig1}b). Despite all coupling beams being vertically polarized, the effective susceptibility is created for both linear polarization components of the probe. The susceptibility  $\chi(x,y)$ is approximately five times stronger~\cite{Zhang2020} for the cross-polarized component $E_x$ of the probe than for the co-polarized one $E_y$. After passing through the Rb cell, only the cross-polarized component of the probe $E_x$ is selected by the polarizing beam splitter (PBS, Fig.~\ref{fig1}a). If the OAM of the transmitted probe is non-zero, one can observe a density dip corresponding to a phase singularity (OAM quantum) just after the PBS (Fig.~\ref{fig1}c). By interfering the output probe with a reference beam originating from the same laser source, we observe interference fringes with a dislocation (Fig.~\ref{fig1}d). The dislocation marks the vortex position and allows us to determine the sign of the generated OAM.

\begin{figure}[tbp]
\centering
\includegraphics[width=0.8\linewidth]{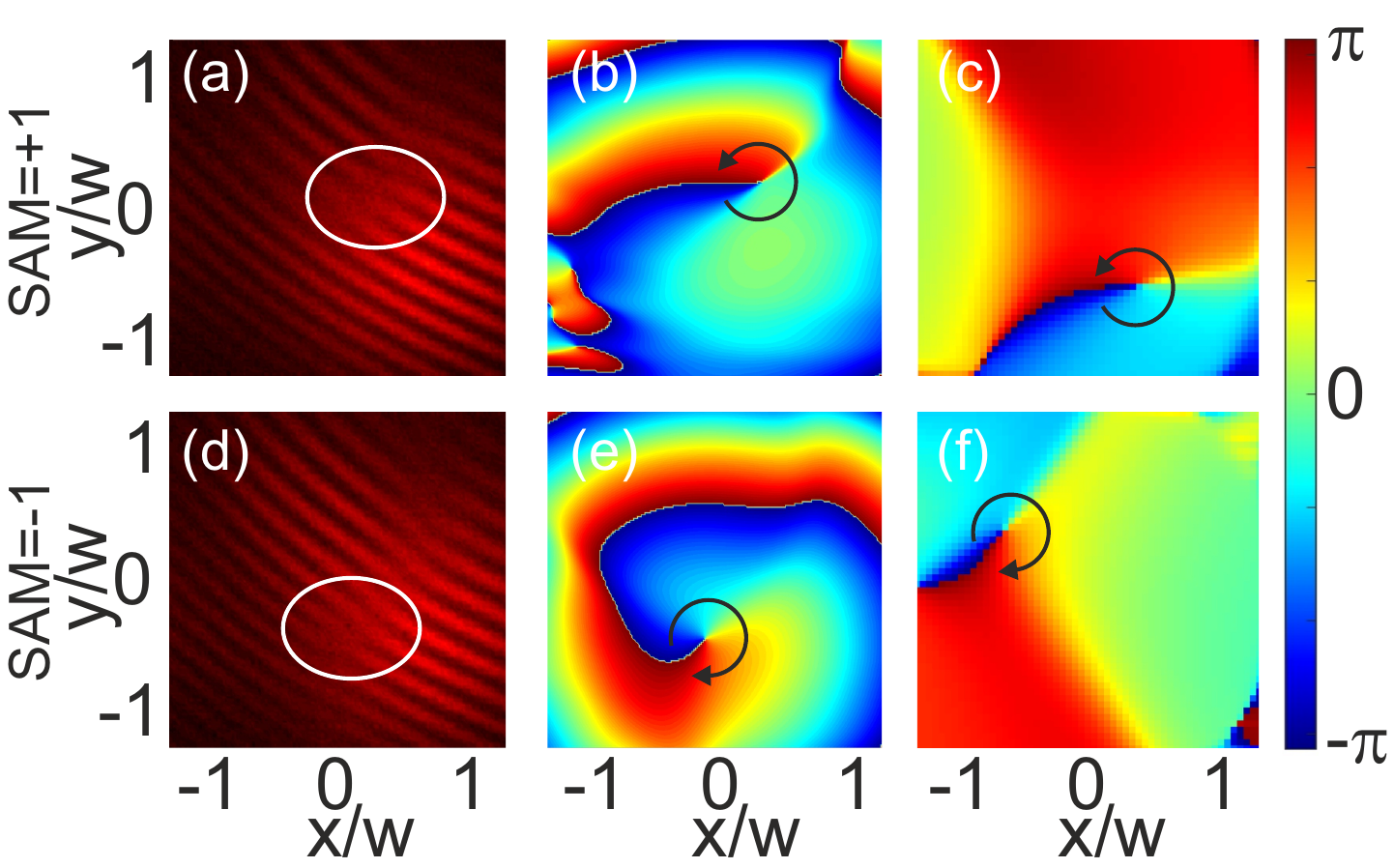}
\caption{\small  \textbf{OAM control by SAM: experiment and theory.} Top row \textbf{(a,b,c)}: left-circular probe; bottom row \textbf{(d,e,f)}: right-circular probe. The staggering is positive: $\Delta>0$. \textbf{(a,d)} Interference pattern (experiment); \textbf{(b,e)} Electric-field phase extracted from (a,d) and exhibiting opposite vortices marked by black arrows (experiment); \textbf{(c,f)} Electric-field phase (theory).
}
\label{fig3}
\end{figure}

In the following, we first present the experimental results demonstrating SAM to OAM conversion and its control, and then discuss the mechanism behind these effects.
Figure~\ref{fig3} shows the SAM to OAM conversion for different circular polarization (SAM) of the probe beam. The top row of the figure shows the results for a left-circular probe, while the bottom row corresponds to a right-circular probe. Experimentally measured interference images demonstrate the resulting OAM  $l=+1$ (panel a, one vortex) and $l=-1$ (panel d, one anti-vortex). This is confirmed by the spatial distribution of the phase (panels b,e) extracted from the experimentally measured interference (see Methods). The positions of the vortices are marked with circles.  Panels (c,f) show theoretical results, as we describe later.

\begin{figure}[tbp]
\centering
\includegraphics[width=0.8\linewidth]{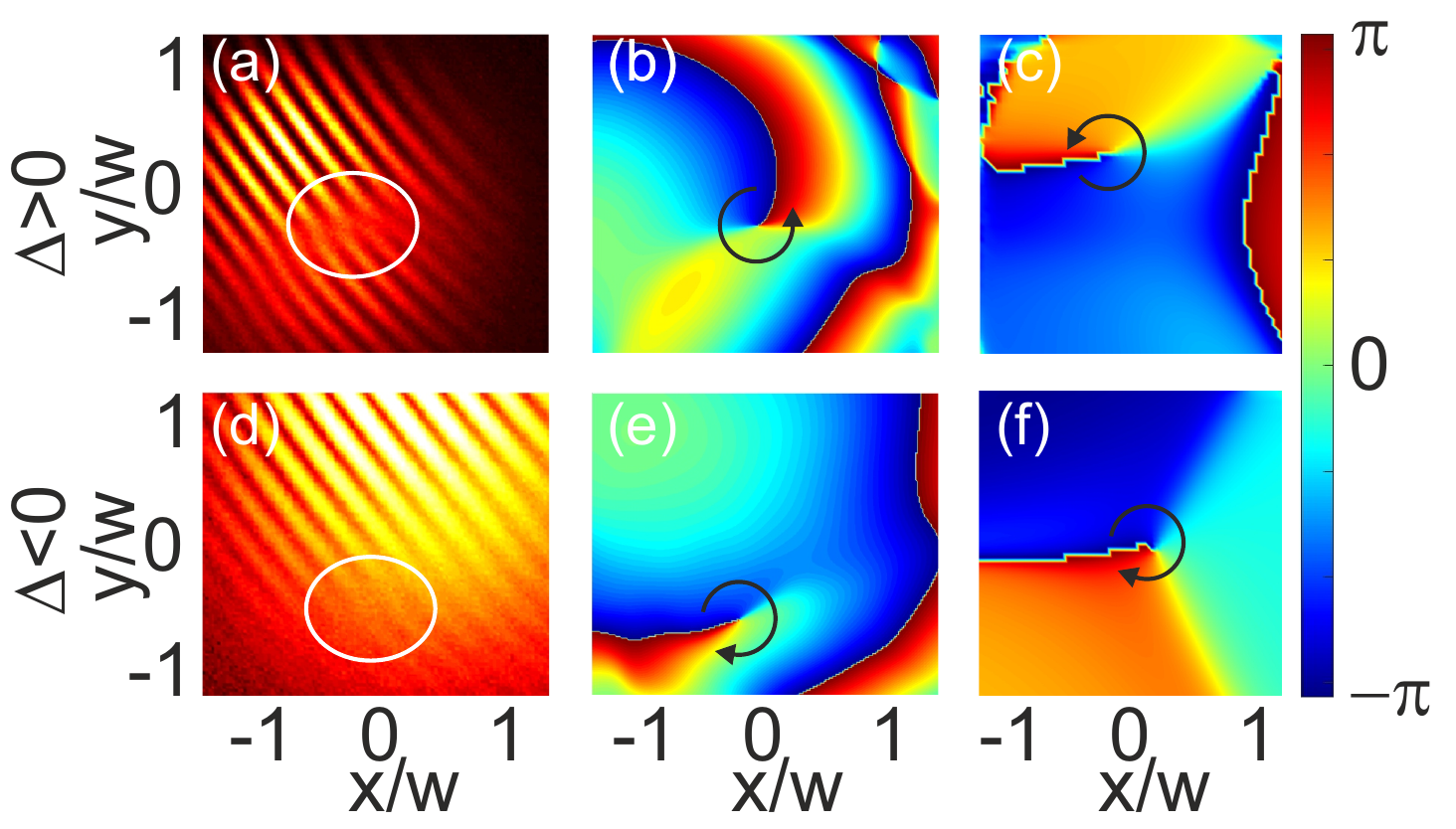}
\caption{\small \textbf{OAM control by staggering: experiment and theory.} 
The incident probe beam is left-circularly polarized. Top row (\textbf{a,b,c}): positive staggering; bottom row (\textbf{d,e,f}): negative staggering.
\textbf{(a,d)} Interference pattern (experiment); \textbf{(b,e)} Electric-field phase extracted from (a,d) and exhibiting opposite vortices marked by black arrows (experiment); \textbf{(c,f)} Electric-field phase (theory).
}
\label{fig2}
\end{figure}

Figure~\ref{fig2} demonstrates the OAM control by the lattice staggering. As in Fig.~\ref{fig3}, we present experimentally measured interference patterns (a,d) and extracted phase (b,e). For a positive staggering, the output OAM is $l=+1$ (top row), whereas for a negative staggering, the output OAM is $l=-1$ (bottom row). Panels (c,f) present the results of numerical simulations discussed below.

The results of experiments in both configurations are summarized in the table~\ref{table1}. In both cases (Figs.~\ref{fig3} and \ref{fig2}), the resulting non-zero OAM is ensured by the SOC present in photonic graphene. The mechanism of its generation is discussed in details below, using a theoretical model based on the paraxial and slowly-varying-envelope approximations.

\begin{table}[H]
\centering
\begin{tabular}{ |c|c|c| } 
\hline
\diagbox{SAM}{Stagg. $\Delta$}&+1&-1\\
\hline
+1&+1&-1\\
-1&-1&+1\\
\hline
\end{tabular}
\caption{\textbf{SAM to OAM conversion and OAM control by the lattice staggering}.}
\label{table1}
\end{table}

The probe beam has a dominant wavevector component $k_0$ along the $z$ axis of the system. Its behavior in the transverse plane can be described in the paraxial approximation by the Schr\"odinger equation, with time $t = z/c$, mass $m = \hbar k_0/c$, and potential $U = - \hbar c k_0 \chi/2$, with $\chi$ being the electric susceptibility and $c$ being the speed of light. The variation $k_z$ of the wavevector component along the $z$ axis plays the role of energy. The presence of two polarizations requires a spinor Schr\"odinger equation with two components, which are uncoupled in the 0th order but coupled in the 2nd order of the small parameter $a/l$ ($a$ is the lattice constant and $l$ is the cell size). The paraxial equations for the 2nd-order components accounting for the SOC read (see Section 2 of Supplementary Materials for details):
\begin{subequations} \label{eq_second_order}
\begin{align}
2 i \frac{\partial E_x^{(2)}}{\partial \zeta} &\approx  - \frac{\partial^2 E_x^{(2)}}{\partial \xi^2} - \frac{\partial^2 E_x^{(2)}}{\partial \eta^2} - \tilde{\chi}_x E_x^{(2)} - \frac{\partial^2  }{\partial \xi \partial \eta}\left(\tilde{\chi}_y E_y^{(0)}\right), \\
2 i \frac{\partial E_y^{(2)}}{\partial \zeta} & \approx - \frac{\partial^2 E_y^{(2)}}{\partial \xi^2} - \frac{\partial^2 E_y^{(2)}}{\partial \eta^2} - \tilde{\chi}_y E_y^{(2)}-\frac{\partial^2  }{\partial \xi \partial \eta}\left(\tilde{\chi}_x E_x^{(0)}\right) .
\end{align}
\end{subequations}
where the rescaled coordinates are $\xi=x/a$, $\eta=y/a$ and $\zeta=z/l$;  $\chi_x$ and $\chi_y$ are the $x$ and $y$ components of the diagonal susceptibility tensor $\chi$, respectively. As shown recently~\cite{Zhang2020}, the TE-TM SOC of light, sometimes called the form birefringence~\cite{alberucci2018photonic}, in the EIT configuration leads to the presence of second-order derivative SOC terms in the linear polarization basis. They include mixed first-order derivatives of both susceptibility and field gradients (transverse wave vectors), contrary to the TE-TM field in cavities, which contains only second-order ones in field gradients (or wave vectors).  
This coupling was already shown to mix the $s$ and $p$ bands of the photonic graphene with consequent OAM generation from linearly polarized probe~\cite{Zhang2020}. However, the realization of SAM to OAM conversion due to this term has not been shown so far.

The strength of the $s$-$p$ band mixing by the SOC depends on the proximity of these bands. To enhance it, we reduce the strength of the potential of photonic graphene.
This potential becomes weak compared to the recoil energy ($\max(|U|)\ll \hbar^2\pi^2/2ma^2$, where $a$ is the lattice period). In this case the nearly free particle approximation~\cite{ablowitz2012nonlinear} is much more suitable than the tight-binding one. We use three slowly varying components A, B, C (centered on A and B sites and the hexagon center, respectively). They describe the three lowest bands strongly mixed at the Dirac point: the two branches of the $s$-band and one of the branches of the $p$-band (See Fig.~S2). We take into account two projections of polarization and two leading orders (0th and 2nd), and thus we construct a $12\times 12$ Hamiltonian $H_{12}$ for slowly varying envelopes, from which we show below a $4 \times 4$ subspace for the states of interest $A_x^{(2),C},A_y^{(0),A},A_y^{(0),B},A_y^{(0),C}$  (see Sections 3-5 of Supplementary Materials):
\begin{equation}
H_{4} = \left( {\begin{array}{*{20}{c | c c c}}
{ - i{\gamma _{xC}}}&{it}&{ - it}&0\\ \hline
{ - it}&{ \Delta- i{\gamma _{yA}}}&{\delta \left( {{q_x} + i{q_y}} \right)}&{\delta \left( {{q_x} - i{q_y}} \right)}\\
{it}&{ \delta \left( {{q_x} - i{q_y}} \right)}&{-\Delta - i{\gamma _{yB}}}&{\delta \left( {{q_x} + i{q_y}} \right)}\\
0&{ \delta \left( {{q_x} + i{q_y}} \right)}&{ \delta \left( {{q_x} - i{q_y}} \right)}&{E_C - i{\gamma _{yC}}}
\end{array}} \right)
\label{Hamiltonian_4}
\end{equation}
where $\delta$ is the coefficient of the Dirac-like terms coupling pairs of modes, $\textbf{q}=(q_x,q_y)$ is the  wavevector counted from the $K$-point, $\Delta$ is the staggering (difference of the average potential energy of the states A and B), $\gamma_{x/y,A/B/C}$ are the decay rates, and $t$ are the SOC coefficients (which depend on $\Delta$) responsible for the conversion between $x$ and $y$ polarizations. 
The Hamiltonian~\eqref{Hamiltonian_4} includes three $y$-polarized bands (the bottom-right block) with Dirac-like couplings, each exhibiting a winding, which can give rise to OAM generation~\cite{Chang2008,song2015unveiling,Zhang2019}. They are spin-orbit coupled (the antidiagonal blocks) to a single $x$-polarized band (the top-left block), where the observations of the probe evolution are performed. 
The advantage of this simplified model is that it allows to understand qualitatively the role of the mechanisms involved in the SAM to OAM conversion.

First, let us consider the circular polarization (SAM) of the beam as a control knob, that is, SAM to OAM conversion in Fig.~\ref{fig3}. The Gaussian probe exhibits zero OAM in all three components (A,B,C). Due to the difference in the decay rates (that is, the non-Hermiticity) $\gamma_{A/B}\gg\gamma_C$, which occurs because the EIT is stronger at the hexagons and the absorption is suppressed, the $C$ component quickly becomes dominant (in both $x$ and $y$, but we first focus on the longer-living $y$-component). The Dirac couplings $\delta$ create opposite OAM quanta in A and B components out of the zero-OAM C component. This is the first key effect of non-Hermiticity. The SOC terms $t$ couple these A and B components to the C component of the other polarization. The OAM quanta generated in $A_y^{(0),A},A_y^{(0),B}$ are converted into $A_x^{(2),C}$. The $x$ polarization (where the observations are made) decays faster than the $y$ polarization ($\gamma_x\gg \gamma_y$), so the 2nd-order signal $A_x^{(2),C}$ becomes comparable with the 0th-order $A_x^{(0),C}$ due to the significant decay of the latter. This is the second key effect of non-Hermiticity.
Due to the opposite signs of the SOC terms, one SOC-transferred OAM quantum in $A_x^{(2),C}$ interferes constructively with the zero-OAM signal in $A_x^{(0),C}$, while the other interferes destructively, resulting in a single vortex (anti-vortex) in the detection. It is the destructive interference which favors the observation of the vortex. Flipping the SAM of the probe changes the relative phase between $x$ and $y$ projections and therefore flips the condition for destructive interference. The simulation results are shown in Fig.~\ref{fig3}c,f for left and right circularly polarized probes, respectively. The change in the SAM of the probe switches the detected OAM from $l = +1$ (Fig.~\ref{fig3}c) to $l = -1$ (Fig.~\ref{fig3}f). The simulation perfectly captures the SAM to OAM conversion observed in the experiment (Fig.~\ref{fig3}b,e).

Second, we consider the case of staggering $\Delta$ as a control knob. One crucial feature of the Hamiltonian Eq.~\eqref{Hamiltonian_4} is the linear dependence of the SOC terms $t$ on the staggering via the  mixed derivatives of the type $\partial_x \chi_{x,y} \partial_y E_{x,y}$ (see Eq.~\eqref{eq_second_order} and Sections 3-5 of Supplementary Materials). The staggering induces a variation of $\chi$, and the flipping of the staggering inverts the sign of the SOC terms $t$. The conditions for constructive and destructive interference are therefore inverted, and so is the sign of the detected OAM quantum.
The results of the numerical simulation are shown in Fig.~\ref{fig2}c,f for positive and negative staggering $\Delta$, respectively. The initial circularly polarized probe shows, after detection in $x$ polarization, OAM $l=+1$ for positive staggering and OAM $l=-1$ for negative staggering. 
Once again, the simulation is in full correspondence with the OAM generation observed in the experiment (Fig.~\ref{fig2}b,e). The use of staggering allows to invert the output OAM for right or left input SAM (see Table~\ref{table1}).

\section*{Discussion}

We have presented an approach for SAM to OAM conversion from an unstructured beam based on a photonic graphene lattice. Our experiment is based on EIT in Rb vapor cell. Its essential ingredients are the non-zero Berry curvature of graphene bands at the Dirac point, the non-Hermiticity, and the photonic SOC. The resulting OAM sign can be flipped by the circular polarization of the incident beam (SAM) or by the lattice staggering, providing efficient and controllable SAM to OAM conversion.

This work underlines the significant interplay of two ingredients common for both control regimes of SAM to OAM conversion. First, the pairs of $y$-polarized modes form Dirac Hamiltonians.
The link of the Dirac Hamiltonian to the formation of OAM quanta (vortex and anti-vortex for different bands) was reported and understood quite recently~\cite{Chang2008,song2015unveiling,Zhang2019}. 
Here, it also involves the non-Hermiticity responsible for the formation of two opposite OAM quanta in two modes from the initial signal with zero OAM.
Second, the SOC provides a selective transfer of one of these quanta to the polarization in which the detection is performed.

It is important to stress that this method is not limited to a particular implementation of photonic graphene. While our implementation with lattice generated by optical beams has the advantage of versatility and reconfigurability, other methods of engineering optical potential can also be used, such as non-linear crystals \cite{heanue1994volume,fleischer2003observation,song2015unveiling}, or liquids \cite{Braidotti2022}, as well as pre-fabricated structures like coupled waveguides \cite{alberucci2018photonic,Rechtsman2013b,petersen2014chiral,abujetas2020spin}, or liquid crystals \cite{belyakov1979optics,Wu2022}. In all these systems the SOC can play an important role for OAM generation and information encoding.
Our method provides the information conversion between probe SAM or lattice staggering, and the OAM, which is particularly promising for applications in optical communications and computing.


\begin{methods}
\textit{Experimental scheme}.
The Gaussian probe beam $\bm{E}_1$ (with a diameter of $\sim0.5$ mm) is derived from an external-cavity diode laser (ECDL1), and its polarization is set as either left or right circular by a quarter wave plate ($\lambda/4$ in Fig.~\ref{fig1}a). The reference beam $\bm{E}_{ref}$ is also from same ECDL1. The 2D hexagonal coupling field (with a period of $\sim43.8$ $\mu$m) is established by interfering three Gaussian beams $\bm{E}_{c1}$, $\bm{E}_{c1}'$ and $\bm{E}_{c1}''$ (with their powers being 6.9 mW, 7.4 mW, and 9.6 mW) from ECDL2, while the 1D coupling field is the interference fringes of Gaussian $\bm{E}_{c2}$ and $\bm{E}_{c2}'$ (with their powers being 4.6 mW and 6.3 mW) from ECDL3. The five $y$-polarized coupling beams possess the same diameter of $\sim1$ mm. The two coupling fields are combined by a beam splitter and then are sent into the vapor cell. The Rb vapor cell with a length of 5 cm is heated to $\sim74^\circ$C to provide an atomic density of $N\approx10^{12}$ cm$^{-3}$. The wavelength of all the involved laser beams is around 795 nm.

Both 1D periodic coupling field and hexagonal coupling field generate EIT windows on the probe field. The susceptibility experienced by the probe field $\bm{E}_1$ is given as:
\begin{equation}
    \chi=\frac{iN\left|\mu_{31}\right|^{2}}{\hbar\varepsilon_{0}}\times\frac{1}{\left(\Gamma_{31}-i\delta_{1}\right)+\frac{\left|\Omega_{2}\right|^{2}}{\Gamma_{32}-i\left(\delta_{1}-\delta_{c1}\right)}+ \frac{\left|\Omega_{3}\right|^{2}}{\Gamma_{32}-i\left(\delta_{1}-\delta_{c2}\right)}}
    \label{EIT}
\end{equation}
in which $N$, $\hbar$, and $\varepsilon_{0}$ are the atomic density, the reduced Planck constant, and the vacuum dielectric constant, respectively; $\delta_i$ is the frequency detuning of field $\bm{E}_i$ ($i$=1, c1 and c2); $\mu_{31}$ and $\Gamma_{31}$ are the dipole moment and decay rate, respectively, between the energy levels $\ket{3}$ and $\ket{1}$ connected by the probe beam (see the inset in Fig.~\ref{fig1}a).

\textit{Data treatment}. To extract the phase from the interference pattern, we first apply a 2D Fourier transform to the measured pattern. We then select one of the two maxima corresponding to the period of intereference and filter out everything else. The remaining maximum is then shifted to $k=0$ and Fourier-transformed back to real space, after which the phase of the resulting complex function is plotted~\cite{Sala2015}.

\textit{Numerical simulations}. We solve the time-dependent equation with the $12\times 12$ Hamiltonian $H_{12}$ defined in Section 4 of Supplementary Materials numerically, by first evaluating each of the matrix elements forming the matrix $H_{12}$ numerically based on Eq.~\eqref{EIT} for $\chi$ (this involves finding the eigenstates A, B, C of the potential described by $\chi$), and then by calculating the time evolution using the 4th order Runge-Kutta method with the Graphics Processing Unit acceleration.

\end{methods}

\bibliographystyle{naturemag}
\bibliography{biblio}

\begin{addendum}
\item We thank Soufiane Hajji for useful discussions. This work was supported by National Natural Science Foundation of China (No.52488201, No.62475209). We also acknowledge the support of the European Union's Horizon 2020 program, through a FET Open research and innovation action under the grant agreement No. 964770 (TopoLight), project ANR Labex GaNEXT (ANR-11-LABX-0014), the ANR project MoirePlusPlus (ANR-23-CE09-0033), and the ANR project "NEWAVE" (ANR-21-CE24-0019), and of the ANR program "Investissements d'Avenir" through the IDEX-ISITE initiative 16-IDEX-0001 (CAP 20-25). 
\item[Author contributions]
Z. Zhang -- project administration, investigation, formal analysis, funding acquisition, methodlogy, visualization, investigation, writing; Y. Shen- investigation, F. Liu-formal analysis, Z. liu-formal analysis, M. Liu-funding acquisition, Y. Zhang-funding acquisition, M. Xiao-supervision; P.~Kokhanchik -- conceptualization,  formal analysis, methodology, visualization, writing; D.~Solnyshkov -- conceptualization, funding acquisition, formal analysis, methodology, visualization, writing; G.~Malpuech -- conceptualization, funding acquisition, methodology, writing, supervision.
 \item[Competing Interests] The authors declare that they have no
competing financial interests.
 \item[Correspondence] Correspondence
should be addressed to : zhyzhang@xjtu.edu.cn (Z.Z);  guillaume.malpuech@uca.fr (G.M.); dmitry.solnyshkov@uca.fr (D.S.).
\end{addendum}

\section*{Data availability statement}
The datasets generated during and/or analysed during the current study are available from the authors upon reasonable request.

\clearpage

\section*{Supplementary Materials}

Below, we present the Supplementary Materials for our manuscript.

\renewcommand{\thefigure}{S\arabic{figure}}
\setcounter{figure}{0}
\renewcommand{\theequation}{S\arabic{equation}}
\setcounter{equation}{0}

\section{Structure of photonic potential}

\begin{figure}[tbp]
\centering
\includegraphics[width=0.95\linewidth]{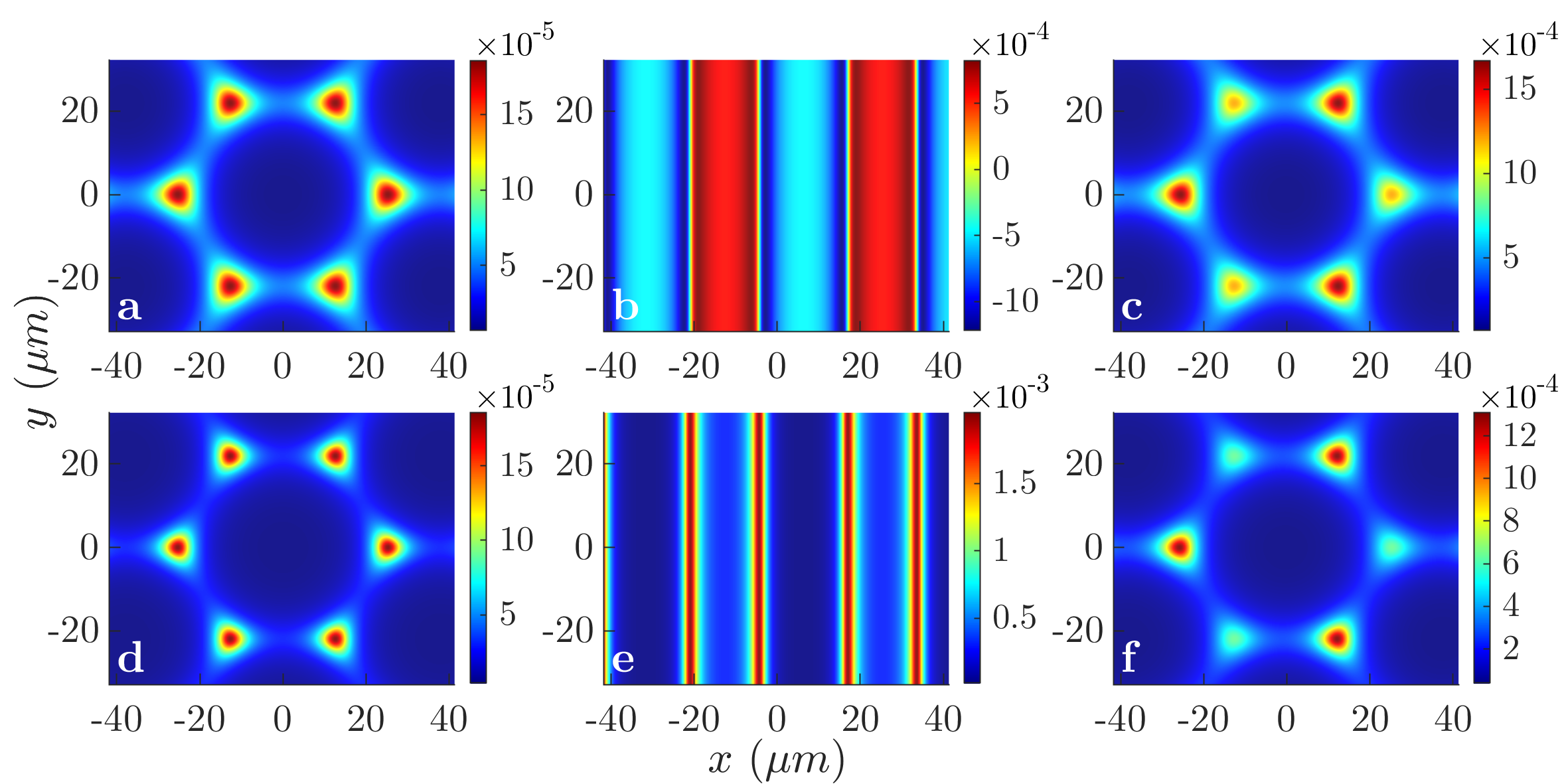}
\caption{\textbf{(a-c)} Real and \textbf{(d-f)} imaginary parts of \textbf{(a,d)} honeycomb, \textbf{(b,e)} staggering, and \textbf{(c,f)} staggered honeycomb susceptibility for X polarization $\chi_x$.}
\label{fig_sm_1}
\end{figure}

The photonic potential $V$ is created by two sets of Y-polarized coupling lasers as described in Methods. The potential $V_x$ for X polarization is simply equal to $-\chi_x$ as we demonstrate in the next section, therefore, we will use the terms 'susceptibility' and 'potential' interchangeably. When two sets of lasers are separated, the first set creates a honeycomb (photonic graphene) potential (Fig.~\ref{fig_sm_1}(a,d)) while the second creates a staggering potential (Fig.~\ref{fig_sm_1}(b,e)) (1D interference fringes). When combined, they give rise to a staggered honeycomb potential (Fig.~\ref{fig_sm_1}(c,f)). The staggering magnitude and sign can be controlled by frequency detuning $\delta_{c2}$ of the $E_{c2}$ and $E_{c2'}$ lasers (see Eq.~(3) in the main text). In Fig.~\ref{fig_sm_1}, we show an example of a single staggering value. The staggering is created for both real (Fig.~\ref{fig_sm_1}(c)) and imaginary (Fig.~\ref{fig_sm_1}(f)) parts of the resulting potential. The potential for the Y polarization aligned along the coupling lasers polarization is approximately 5 times smaller for the real part $Re[\chi_y] \approx 0.2 Re[\chi_x]$ and 10 times smaller for the imaginary part $Im[\chi_y] \approx 0.1 Im[\chi_x]$ in correspondence with our previous experiments~\cite{Zhang2020}.

\section{Paraxial approximation of the wave equation with photonic spin-orbit coupling}

The wave equation for electromagnetic waves in inhomogeneous dielectric media can be written as:
\begin{equation} \label{wave_eq}
    -\Delta \textbf{E} + \nabla (\nabla , \textbf{E}) = -\frac{1 + \chi}{c^2} \frac{\partial^2 \textbf{E}}{\partial t^2},
\end{equation}
where $\textbf{E}$ is the electric field, $\chi$ is the electric susceptibility tensor, $c$ is the speed of light in vacuum, $t$ is time, $\Delta$ and $\nabla$ are the Laplacian and nabla operators, respectively, acting in 3D space $\{x,y,z\}$.

The dynamics of the light beam passing along the $z$ axis through the atomic vapor cell in the EIT regime is well described in the paraxial approximation~\cite{lax1975maxwell}:
\begin{equation} \label{ansatz_paraxial}
    \textbf{E}(x,y,z) = \tilde{\textbf{E}}(x,y,z) e^{i k_0 z - i \omega t},
\end{equation}
\begin{equation} \label{condition_paraxial}
    \left| \frac{\partial^2 \tilde{\textbf{E}}}{\partial z^2} \right| \ll \left| k_0 \frac{\partial \tilde{\textbf{E}}}{\partial z} \right| \ll k_0^2,
\end{equation}
where $k_0 = \omega/c$ is the wavevector along $z$ axis, $\omega$ is the wave frequency. We can introduce characteristic lengths in $(x,y)$ plane as $w$ and along $z$ axis as $l=k_0 w^2$. Then, the paraxial approximation is described by the small parameter $f = w/l$, which is $\sim 10^{-2}$ in our case. We can also introduce dimensionless coordinates $\xi=x/w$, $\eta=y/w$ and $\zeta=z/l$. We notice that the potentials $\chi_{x,y}$ are of the order of $10^{-4}$, therefore, the potentials can be rescaled as $\tilde{\chi}_{x,y} = \chi_{x,y}/f^2$. Finally, using all these notations, Eq.~\eqref{wave_eq} can be rewritten as a following system of equations:
\begin{subequations} \label{paraxial_eq}
\begin{align} 
f \frac{\partial^2 E_y}{\partial \xi \partial \eta} + i \frac{\partial E_z}{\partial \xi} + f^2 \frac{\partial^2 E_z}{\partial \xi \partial \zeta} - f \frac{\partial^2 E_x}{\partial \eta^2} - f^3 \frac{\partial^2 E_x}{\partial \zeta^2} - 2 i f \frac{\partial E_x}{\partial \zeta} & = f \tilde{\chi}_x E_x, \\
f \frac{\partial^2 E_x}{\partial \xi \partial \eta} + i \frac{\partial E_z}{\partial \eta} + f^2 \frac{\partial^2 E_z}{\partial \eta \partial \zeta} - f \frac{\partial^2 E_y}{\partial \xi^2} - f^3 \frac{\partial^2 E_y}{\partial \zeta^2} - 2 i f \frac{\partial E_y}{\partial \zeta} & = f \tilde{\chi}_y E_y, \\
f^2 \frac{\partial^2 E_x}{\partial \xi \partial \zeta} + i \frac{\partial E_x}{\partial \xi} + f^2 \frac{\partial^2 E_y}{\partial \eta \partial \zeta} + i \frac{\partial E_y}{\partial \eta} - f \frac{\partial^2 E_z}{\partial \xi^2} - f \frac{\partial^2 E_z}{\partial \eta^2} & = f^{-1} E_z.
\end{align}
\end{subequations}

Now, we can write a series decomposition of the fields with respect to the small parameter $f$. It turns out, that the $E_{x,y}$ fields contain only even terms of the decomposition, while $E_z$ contains only odd terms:
\begin{subequations} \label{eq_decompositions}
\begin{align}
E_{x,y} &= E_{x,y}^{(0)} + f^2 E_{x,y}^{(2)} + ... \\
E_z &= f E_{z}^{(1)} + f^3 E_{z}^{(3)} + ...
\end{align}  
\end{subequations}
Zero-order ($f^0$) and first-order ($f^1$) terms in Eq.~\eqref{paraxial_eq} provide us with a typical system of equations describing dynamics of zero-order fields $E_{x,y}^{(0)}$ in the paraxial approximation:
\begin{subequations} \label{eq_zero_order}
\begin{align}
2 i \frac{\partial E_x^{(0)}}{\partial \zeta} &= - \frac{\partial^2 E_x^{(0)}}{\partial \xi^2} - \frac{\partial^2 E_x^{(0)}}{\partial \eta^2} - \tilde{\chi}_x E_x^{(0)}, \\
2 i \frac{\partial E_y^{(0)}}{\partial \zeta} &= - \frac{\partial^2 E_y^{(0)}}{\partial \xi^2} - \frac{\partial^2 E_y^{(0)}}{\partial \eta^2} - \tilde{\chi}_y E_y^{(0)}.
\end{align}  
\end{subequations}
Eq.~\eqref{eq_zero_order} is a Schrödinger-like equation with the time coordinate replaced by the real space coordinate $\zeta$. The first and the second terms on the RHS of Eq.~\eqref{eq_zero_order} stand for kinetic energy, while the third term is the photonic potential induced by coupling beams (see the previous section of Supplementary Materials). As one can see, these equations do not involve spin-orbit coupling terms and two polarizations evolve independently. Therefore, these equations are not enough to explain the SAM to OAM conversion controlled by the excitation polarization (Fig.~2 of the main text). So, we have to go to the next order to include the polarization coupling. $f^2$ and $f^3$ terms of Eq.~\eqref{paraxial_eq} give us the dynamics of the second-order fields $E_{x,y}^{(2)}$:
\begin{subequations} \label{eq_second_order}
\begin{align}
2 i \frac{\partial E_x^{(2)}}{\partial \zeta} &= - \frac{\partial^2 E_x^{(2)}}{\partial \xi^2} - \frac{\partial^2 E_x^{(2)}}{\partial \eta^2} - \tilde{\chi}_x E_x^{(2)} + S_{xx}[\tilde{\chi}_x,E_x^{(0)}] + S_{xy}[\tilde{\chi}_y E_y^{(0)}], \\
2 i \frac{\partial E_y^{(2)}}{\partial \zeta} &= - \frac{\partial^2 E_y^{(2)}}{\partial \xi^2} - \frac{\partial^2 E_y^{(2)}}{\partial \eta^2} - \tilde{\chi}_y E_y^{(2)} + S_{yy}[\tilde{\chi}_y,E_y^{(0)}] + S_{yx}[\tilde{\chi}_x E_x^{(0)}].
\end{align}
\end{subequations}
As one can see, the dynamics of the 2nd-order field components is similar to the one in Eq.~\eqref{eq_zero_order} with the addition of contribution from the 0th-order through $S_{\alpha \beta}$ functionals. Since we expect that the SOC plays the central role in the SAM to OAM conversion, we keep only cross-polarized terms $S_{xy}[g] = S_{yx}[g] = - \frac{\partial^2 g}{\partial \xi \partial \eta}$.

\section{Spectrum of the system}

It is possible now to straightforwardly solve Eqs.~\eqref{eq_zero_order} and~\eqref{eq_second_order} for dynamics. These simulations can reproduce the experimental results but are not strongly helpful for understanding the mechanism behind the SAM to OAM conversion. Therefore, we would like to apply slowly varying envelope approximation (SVEA) to Eqs.~\eqref{eq_zero_order} and~\eqref{eq_second_order} in order to build a simpler model, both from simulation and analysis points of view.

\begin{figure}[tbp]
\centering
\includegraphics[width=0.7\linewidth]{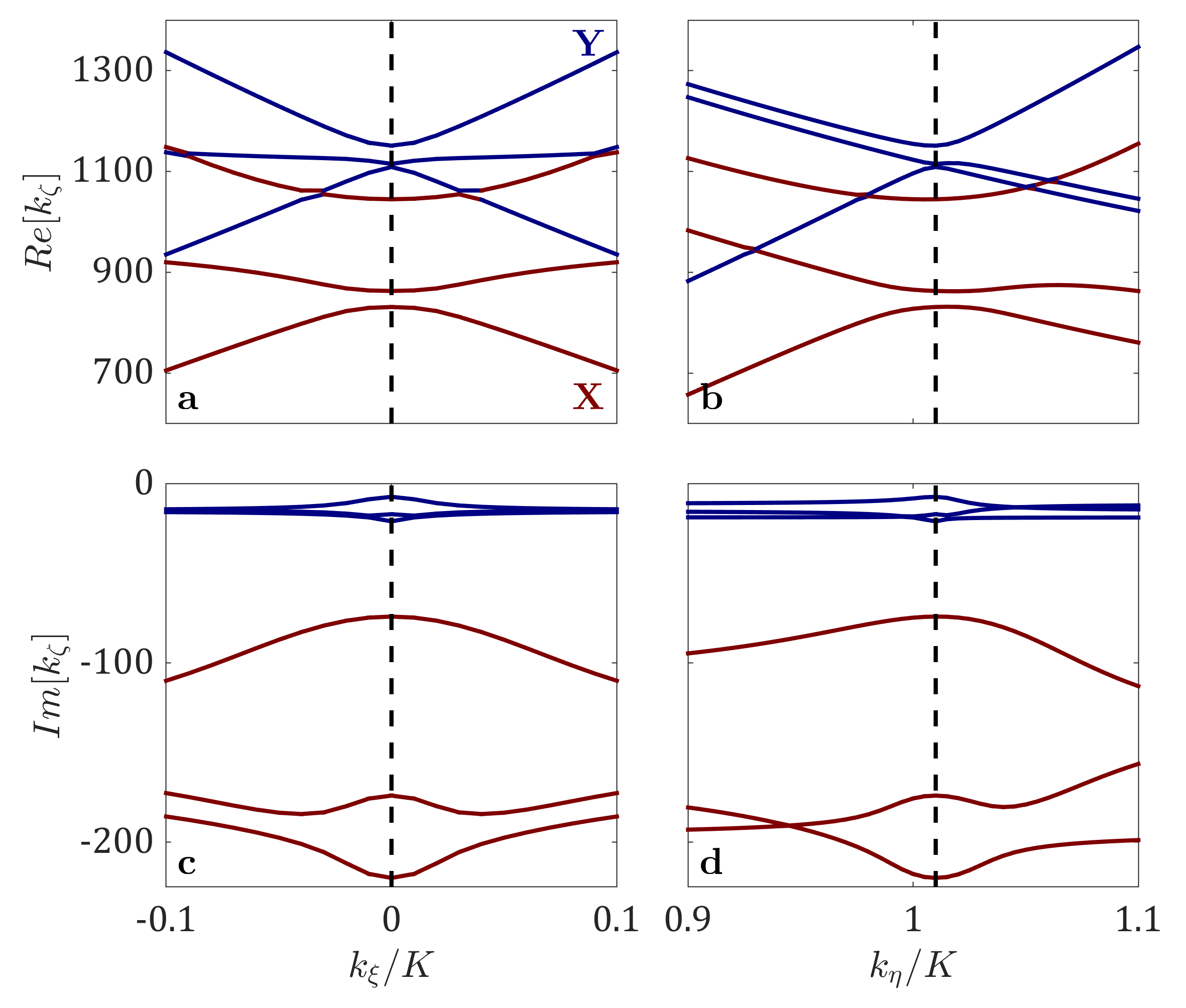}
\caption{Dispersion of 'lower-energy' states of Eq.~\eqref{eq_zero_order} along $k_\xi$ \textbf{(a,c)} and $k_\eta$ \textbf{(b,d)}; upper row \textbf{(a,b)} -- real part of $k_\zeta$, lower row \textbf{(c,d)} -- imaginary part of $k_\zeta$; red (blue) color corresponds to the X (Y) polarization; dashed black line shows a point in $k$ space for which we plot eigenmodes in Fig.~\ref{sm_fig_eigenmodes}.}
\label{sm_fig_dispersion}
\end{figure}

\begin{figure}[tbp]
\centering
\includegraphics[width=0.95\linewidth]{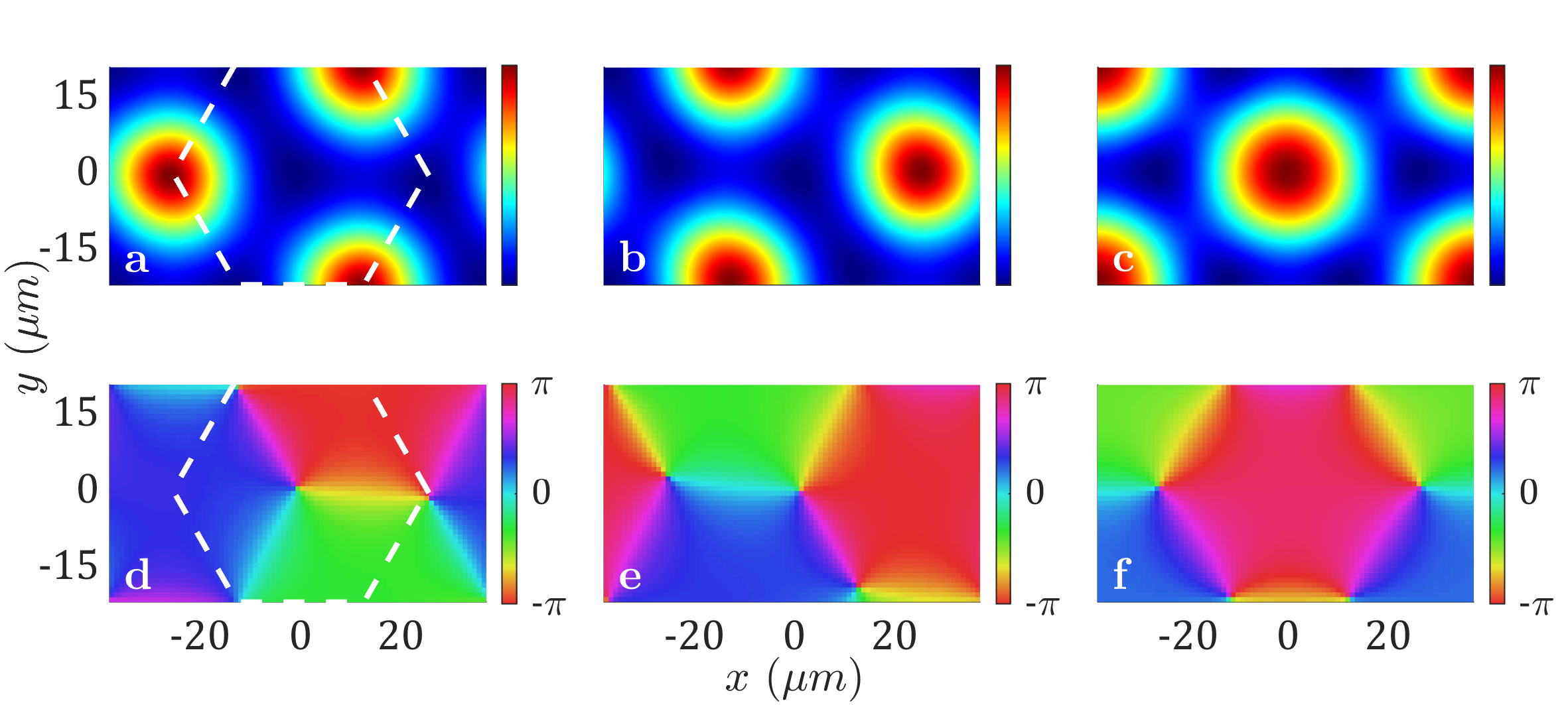}
\caption{Three 'lower-energy' eigenmodes in X polarization near the $K$ point of the Brillouin zone: \textbf{(a-c)} real space normalized density and \textbf{(d-f)} real space phase; columns are arranged in order of ascending $k_\zeta$ (see black dashed line in Fig.~\ref{sm_fig_dispersion}): left -- A state, middle -- B state, right -- C state; white dashed lines indicate the honeycomb lattice cell.}
\label{sm_fig_eigenmodes}
\end{figure}

We first find the eigenmodes and dispersion of Eq.~\eqref{eq_zero_order}. We perform our analysis in the vicinity of the K point to match the excitation conditions of the experiment. Dispersion of several 'lower-energy' (actually, lower-$k_{\zeta}$, since the role of time is played by the spatial coordinate $\zeta$) is shown in Fig.~\ref{sm_fig_dispersion}. Contrary to the case of honeycomb lattice in the tight-binding approximation, where the set of lower-energy states is formed by two eigenmodes, localized on A and B sites, respectively, here the potential is weak, and another extreme is achieved, namely, nearly free particle approximation. In this case, an additional eigenmode localized on the center (C) of the hexagon is 'energetically' close to the A and B states~\cite{ablowitz2012nonlinear}. The A,B and C eigenmodes for X polarization are shown in Fig.~\ref{sm_fig_eigenmodes}. Since potentials have different strengths for two polarizations, there are in total 6 'lower-energy' eigenmodes which we take into account in the following. The structure of eigenmodes in Y polarization is very similar to the one shown for X polarization. We expect that all important dynamics is happening between these 6 states.

\section{Slowly varying envelope approximation (SVEA)}

Now, we can decompose both zeroth- and second-order fields on the basis of 'lower-energy' states $U_{x,y}^{Z}$ ($Z=\{A,B,C\}$) with slowly varying envelopes $A_{x,y}^{(0,2),Z}$:
\begin{align} \label{eq_svea_decomposition}
    E_{x,y}^{(0,2)}(\xi,\eta,\zeta) &= \sum_Z A_{x,y}^{(0,2),Z}(\xi,\eta,\zeta) U_{x,y}^{Z}(\xi,\eta).
\end{align}
In total, there are 12 slowly varying envelopes $A=\{A_{x,y}^{(0,2),Z}, Z=A,B,C\}$ which define the dimensionality for the 'Hamiltonian' matrix based on Eqs.~\eqref{eq_zero_order} and \eqref{eq_second_order}:
\begin{equation} \label{eq_Ham_matrix}
    2i \frac{\partial A}{\partial \zeta} = \hat{H}_{12} A, \quad \hat{H}_{12} = \hat{K} + \hat{P} + \hat{S},
\end{equation}
where $\hat{K}$ is the kinetic energy matrix, $\hat{P}$ is the potential matrix, and $\hat{S}$ is the SOC matrix. Their non-zero matrix elements are defined as:
\begin{subequations} \label{eq_Ham_components}
\begin{align}
K_{\alpha\alpha,ZZ'} &= {U_{\alpha}^{Z'}}^* \left( - \frac{\partial^2}{\partial \xi^2} - \frac{\partial^2}{\partial \eta^2} \right) U_{\alpha}^{Z}, \\
P_{\alpha\alpha,ZZ} &= - {U_{\alpha}^{Z}}^* \tilde{\chi}_{\alpha} U_{\alpha}^{Z}, \\
S_{\alpha\bar{\alpha},ZZ'} &= {U_{\bar{\alpha}}^{Z'}}^* S_{\bar{\alpha}\alpha}[\tilde{\chi}_\alpha U_\alpha^Z] = -{U_{\bar{\alpha}}^{Z'}}^* \frac{\partial^2}{\partial \xi \partial \eta} \tilde{\chi}_\alpha U_\alpha^Z, \label{eq_matrix_elems_soc}
\end{align}
\end{subequations}
where $\bar{\alpha}$ stands for an orthogonal to $\alpha$ in-plane coordinate (e.g., if $\alpha=x$ then $\bar{\alpha}=y$).

\section{System dynamics}
The exact kinetic energy matrix Eq.~\eqref{eq_Ham_components} can be well approximated by a block diagonal matrix $\hat{K}=\text{diag(} 4\times\{\hat{K}_0\} \text{)}$, where all 4 3$\times$3 blocks $\hat{K}_0$ (one block per field component $E_{x,y}^{(0,2)}$) are identical and equal:
\begin{equation} \label{eq_kin_energy_block}
    \hat{K_0} = \delta
    \begin{pmatrix}
        0 & k_\xi + i k_\eta & k_\xi - i k_\eta \\
        k_\xi - i k_\eta & 0 & k_\xi + i k_\eta \\
        k_\xi + i k_\eta & k_\xi - i k_\eta & 0
    \end{pmatrix}.
\end{equation}
The small parameter $\delta$, basically, defines the slow space coordinates as $(\xi',\eta')=(\xi,\eta)/\delta$. It is worth noting that these kinetic energy blocks are identical for both polarizations since the leading order of the nearly free particle approximation is potential-independent. As one can see, all three basis states of the matrix $\hat{K}_0$ enter symmetrically in the matrix (up to the winding sign of coupling term). Each pair forms a Dirac-like Hamiltonian. Therefore, if a single eigenmode $U_\alpha^Z$ is excited, the vortices of different signs appear in two other eigenmodes (this is the conical refraction~\cite{Zhang2019}). However, if two eigenmodes are excited equally, they create two vortices of opposite signs in the third eigenmode that cancel each other. A similar situation happens when all three eigenmodes are excited equally.

In our experiment, we excite the system with a broad Gaussian beam. This excitation corresponds to a symmetric linear superposition of all three eigenmodes which we use as an initial state $E_\alpha^{(0)}(\zeta=0)=\frac{1}{\sqrt{3}} \sum_Z U_\alpha^Z$. Therefore, the kinetic energy term Eq.~\eqref{eq_kin_energy_block} itself is not enough to create vortices in any polarization. Here, the imaginary part of the potential $\chi_{x,y}$ becomes important. A,B and C states have different decays, with the C state being most long-living between the three. Therefore, after some 'time' of evolution (at some distance $\zeta$), the C mode becomes more populated than the A and B modes, and the situation becomes equivalent to the excitation of the kinetic energy Hamiltonian Eq.~\eqref{eq_kin_energy_block} with a single C component. As a result, two vortices of different signs are created in A and B eigenmodes, one for each eigenmode.

Everything described above happens in the 0th order of a single Y polarization, namely, $E_y^{(0)}$. However, the detection is performed in X polarization, and the detected signal consists of the sum of 0th and 2nd order field components (see Eq.~\eqref{eq_decompositions}). Again, the C state is the most long-living in X polarization as well, and, therefore, the components $A_x^{(0,2),C}$ make the principle contribution in the detection. 0th order component in X $A_x^{(0),C}$ evolves similarly to the corresponding component in Y $A_y^{(0),C}$, namely, it does not have a vortex, as described above. However it shows a quicker decay with respect to $A_y^{(0),C}$ because of a bigger imaginary potential in X (see Fig.~\ref{sm_fig_dispersion}(c,d)). 2nd order component in X $A_x^{(2),C}$ is formed fully due to the spin-orbit coupling term $\hat{S}$ transferring two vortices of opposite signs from $E_y^{(0),A}$ and $E_y^{(0),B}$, respectively. Consequently, the 'lifetime' of $A_x^{(2),C}$ component is defined by the 'lifetime' of Y polarization, which is significantly bigger than the one for X. As a result, despite the definition of the orders in Eq.~\eqref{eq_decompositions}, because of the non-Hermitian nature of the potential, 0th order in X $A_x^{(0),C}$ decays faster than 2nd order in X $A_x^{(2),C}$, and at some moment of 'time' they become comparable. According to our simulations, the length of vapor cell corresponds to this characteristic 'time' (characteristic length along $\zeta$). Therefore, the total (detected) signal is the result of the interference of the constant-phase field $A_x^{(0),C}$ and the vortex-antivortex field $A_x^{(2),C}$.

One of these two vortices interferes constructively with $A_x^{(0),C}$, while another interferes destructively. Whether the interference is constructive or destructive is defined by the structure of spin-orbit coupling matrix elements $S_{yx,AC}$ and $S_{yx,BC}$. These matrix elements are staggering-dependent (since $\tilde{\chi}_\alpha$ enters the Eq.~\eqref{eq_matrix_elems_soc}). That is why by controlling staggering, we can 'choose' which vortex interferes constructively (Fig.~3 of the main text).

The result of interference also depends on the initial phase difference between X and Y polarizations, in other words, on the chirality of the excitation beam. This is so because of the phase of $A_x^{(0),C}$ controlled by the initial phase in X polarization, and phase of $A_x^{(2),C}$ controlled by the initial phase in Y polarization. Therefore, inverting the circular polarization allows us to 'choose' which vortex interferes constructively (Fig.~2 of the main text).

\end{document}